\documentstyle[amsfonts,aps,12pt]{revtex}
\begin{document}
\author{F. Dahia\thanks{%
e-mail: fdahia@fisica.ufpb.br} and C. Romero\thanks{%
e-mail: cromero@fisica.ufpb.br}}
\title{Line Sources in Brans-Dicke Theory of Gravity}
\address{Departamento de F\'{\i}sica\\
Universidade Federal da Para\'{\i}ba\\
Cx. Postal 5008, 58059-970, Jo\~{a}o Pessoa, PB, Brazil}
\maketitle

\begin{abstract}
We investigate how the gravitational field generated by line sources can be
characterized in Brans-Dicke theory of gravity. Adapting an approach
previously developed by Israel who solved the same problem in general
relativity we show that in Brans-Dicke theory's case it is possible to work
out the field equations which relate the energy-momentum tensor of the
source to the scalar field, the coupling constant $\omega $ and the
extrinsic curvature of a tube of constant geodesic radius centered on the
line in the limit when the radius shrinks to zero. In this new scenario two
examples are considered and an account of the Gundlach and Ortiz solution is
included$.$ Finally, a brief discussion of how to treat thin shells in
Brans-Dicke theory is given.
\end{abstract}

\newpage

\section{Introduction}

Topological defects such as cosmic strings, domains walls, monopoles and
textures \cite{vilenkin-livro} have become recently a very significant part
of current theoretical physics research. Predicted by GUT models these
structures are now considered to be viable candidates for explaining a
number of observational phenomena in Astrophysics and Cosmology. Cosmic
strings, in particular, which are thought to have been generated in the
early Universe, during a phase transition, might be responsible for
gravitational lensing \cite{gott} and galaxy formation \cite{brandenberger},
among other physical phenomena. Naturally, from a theoretical standpoint,
the entire subject was first investigated in the context of general
relativity. Recently, however, due to a renewed interest in Brans-Dicke
theory, mainly in connection with the inflation problem, a number of authors 
\cite{gundlach,barros,sen,mex,barros2,sen2,dando,dando2} have started
considering cosmic strings, domain walls and monopoles in scalar-tensor
theories as well as in dilaton gravity. As a result of these investigations
it has been shown that scalar-tensor theories lead to the prediction of new
and different phenomena concerning the gravitational effects produced by
topological defects. As an example, let us mention the appearance of
gravitational forces exerted by a global monopole on the matter around it,
an effect predicted by Brans-Dicke theory and which is absent in the case of
general relativity's monopole \cite{vilenkin-livro,barros2}.

Historically one could say that cosmic strings, the most studied of all
topological defects, entered the cosmological scenario through the work by
Vilenkin \cite{vilenkin} who solved the Einstein field equations for a
matter distribution corresponding to an infinite, static, straight string.
Vilenkin's solution, which was obtained using the weak field approximation
of general relativity depicts the gravitational field generated by the
string as described by a conical space-time whose angular deficit is related
to the mass density. Exact solutions of a simple model of the string were
found later by Gott \cite{gott} and Hiscock \cite{hiscock} independently and
also correspond to conical space-time.

The matter distribution which represents a cosmic string is a very
particular case of an idealization of realist matter distribution which may
be generally called line sources. Long before cosmic strings had become
known in the literature the problem of how to characterize line sources in
general relativity and how to compute the gravitational field generated by
such structures was already discussed by Israel \cite{israel} with a high
degree of generality. As shown by Israel, this is a complex problem and as
yet there is no simple prescription of how to characterize physically an
arbitrary line source. However, for a wide class of line sources one can
work out through the field equations a relation between the line
energy-momentum tensor and the extrinsic curvature of a tube of matter
enclosing the source in the limit when the radius of the tube tends to zero.

The purpose of this paper is to consider the method employed by Israel
concerning line sources in general relativity and extend it to investigate
the same problem in Brans-Dicke theory of gravity. Thus, in section II we
give a general description of the so-called simple line sources in
Brans-Dicke theory. In section III we attempt to assign a ``line
energy-momentum tensor'' density to a simple line. Section IV deals with
applications and includes a discussion of cosmic strings in Brans-Dicke
theory. Finally, a brief account of thin shells or surface layers is given
in section V.

\section{General characterization of line sources in Brans-Dicke theory of
gravity}

Given that lines are essentially idealizations of realistic matter
distributions, we start by partially characterizing a line source as a
``singularity'' that can be enclosed in a tube of arbitrary small radius.
Then, we proceed to give the following definitions:

A two-dimensional submanifold $L$ embedded in space-time is called a
``line'' if the following conditions hold:

(i) There exists a neighborhood $N$ of $L$ such that each point $p\in N$ can
be connected to $L$ by a spacelike curve of finite length and for each $p\in
N$ there exists a minimal length curve, which is a geodesic of length $\rho
(p),$ connecting $p$ to $L.$ This function $\rho (p)$ defines the geodesic
radius of $p$ and may be taken as one of the coordinates of the space-times
points in the neighborhood $N.$ It can be shown that these geodesics are
orthogonal to the hypersurface $\Sigma $ defined by the equation $\rho
=const.$

(ii) The hypersurfaces $\Sigma $ are tridimensional timelike submanifolds
embedded in space-time (``three-cylinders'') with topology $S^1\times
S^1\times {\Bbb R}$ or $S^1\times {\Bbb R}^2$ depending on the lines $L$
being closed or not.

(iii) Each three-cylinder $\Sigma $ may be covered by a congruence of simple
nonreducible closed spacelike curves whose length tends to zero as they are
Lie-transported inward to $L$ along radial geodesics.

Clearly, the definition above naturally suggests the choice of the following
coordinate system to cover the region $N.$ We choose $\rho $ as a radial
coordinate and, taking into account property (iii), choose an angular
coordinate $\varphi $ $\left( 0<\varphi <2\pi \right) $ that acts as a
parametrization of the congruence of curves which covers the family of
hypersurfaces $\Sigma .$ Then we introduce two more coordinates, $z$ and $t,$
in one of the three-cylinders $\Sigma $ and Lie-transport $(z,t)$ along
radial geodesics to cover the entire region $N$. In this way, we end up with
a Gaussian coordinate system in terms of which the metric takes the form 
\begin{equation}
ds^2=d\rho ^2+g_{ij}\left( \rho ,x^k\right) dx^idx^j,  \label{metric-gauss}
\end{equation}
where the Latin indices coordinates $x^i$ denotes $(t,\varphi ,z),$ with $%
t\in {\Bbb R},$ $0<\varphi <2\pi $ and the range of $z$ depends on the
three-cylinders topology.

Now, from the form (\ref{metric-gauss}) of the metric written in the
Gaussian coordinates $(\rho ,t,\varphi ,z)$ we can easily calculate the
extrinsic curvature $K_{ij}$ of the three-cylinders $\Sigma $ along with the
corresponding density ${\cal K}_{ij}$ to obtain 
\begin{equation}
K_{ij}=\frac 12\frac{\partial g_{ij}}{\partial \rho },\qquad {\cal K}_j^i=%
\sqrt{-g}K_j^i.
\end{equation}

On the other hand, the foliation $\rho =const$ allows us to decompose the
Ricci tensor in terms of the extrinsic and intrinsic curvature of the
three-cylinders $\Sigma .$

At this point let us consider Brans-Dicke field equations in the form 
\begin{mathletters}
\label{full-bd}
\begin{eqnarray}
R_{\;\nu }^\mu +W_{\;\nu }^\mu &=&-\frac{8\pi }\phi \left( T_{\;\nu }^\mu -%
\frac 12\delta _{\;\nu }^\mu f(\omega )T\right)  \label{eqs-bd} \\
\Box \phi &=&\frac{8\pi }{3+2\omega }T,  \label{dal-phi}
\end{eqnarray}
where $\phi $ is the scalar field, $\omega $ is the coupling constant, $%
T_{\;\nu }^\mu $ is the energy-momentum tensor of the matters fields, $%
T=T_{\;\mu }^\mu ,$ and we have defined 
\end{mathletters}
\begin{mathletters}
\label{full-bd}
\begin{eqnarray}
W_{\;\nu }^\mu &\equiv &\frac \omega {\phi ^2}\phi ^{,\mu }\phi _{,\nu }+%
\frac 1\phi \phi ^{;\mu }{}_{;\nu }  \label{W} \\
f(\omega ) &\equiv &\frac{2+2\omega }{3+2\omega }.  \label{f(w)}
\end{eqnarray}

One can show that the field equations (\ref{eqs-bd}) and (\ref{dal-phi}) may
be decomposed into 
\end{mathletters}
\begin{mathletters}
\label{bd}
\begin{eqnarray}
\frac \partial {\partial \rho }\left( \phi {\cal K}_{\;j}^i\right) +\phi 
\sqrt{-g}\left( ^{\left( 3\right) }R_{\;j}^i+^{(3)}W_{\;j}^i\right) =-8\pi 
\sqrt{-g{}}\left( T_{\;j}^i-\frac 12\delta _{\;j}^if(\omega )T\right) &&
\label{bd1} \\
K,_j-^{(3)}\nabla _iK_{\;j}^i-\frac{\phi ,_i}\phi K_{\;j}^i+\frac{\phi
,_\rho ,_j}\phi +\omega \frac{\phi _{,\rho }\phi ,_j}{\phi ^2}=-\frac{8\pi }%
\phi T_{\;j}^\rho &&  \label{bd2} \\
K_{ij}K^{ij}-K^2+\omega \left( \frac{\phi ,_\rho }\phi \right) ^2-2K\left( 
\frac{\phi ,_\rho }\phi \right) -^{\left( 3\right) }R-^{(3)}W=-\frac{16\pi }%
\phi T_{\;\rho }^\rho &&  \label{bd3} \\
\frac \partial {\partial \rho }\left( \sqrt{-g}\phi ,_\rho \right) +\sqrt{-g}%
\left( ^{(3)}\nabla ^2\phi \right) =\frac{8\pi }{3+2\omega }\sqrt{-g}T, &&
\label{bd4}
\end{eqnarray}
where $K=K_i^i,$ $^{\left( 3\right) }W=W_i^i,$ $^{\left( 3\right) }R_{\;j}^i$
and the operator $^{(3)}\nabla _i$ denote the Ricci tensor and the covariant
derivative associated with the three-metric $g_{ij},$ respectively.

At this stage let us define the concept of simple line in Brans-Dicke theory:

A line $L$ is called a {\it simple line} if the following additional
conditions hold:

(iv) The intrinsic curvature density of the cylinders $\rho =const$
converges less rapidly than the term $\frac 1{\phi \rho },$ i.e., $%
\lim_{\rho \rightarrow 0}\left( \rho \phi \sqrt{-g}^{\left( 3\right)
}R_{\;j}^i\right) =0.$ This condition, called normal-dominated convergence,
is required to be fulfilled also by $^{\left( 3\right) }W_j^i,$ i.e., $%
\lim_{\rho \rightarrow 0}\left( \rho \phi \sqrt{-g}^{\left( 3\right)
}W_{\;j}^i\right) =0.$ Also, for the term $^{\left( 3\right) }\nabla ^2\phi $
we should have $\lim_{\rho \rightarrow 0}\left( \rho \sqrt{-g}^{\left(
3\right) }\nabla ^2\phi \right) =0.$ (As we shall see, these conditions
imply the existence of the following limits: $\lim_{\rho \rightarrow
0}\left( \phi {\cal K}_{\;j}^i\right) ={\cal C}_{\;j}^i\left( t,\varphi
,z\right) $ and $\lim_{\rho \rightarrow 0}\left( \sqrt{-g}\phi _{,\rho
}\right) =\ell (t,\varphi ,z).$)

(v) We assume that the closed curves of condition (iii) can be chosen in
such a way that 
\end{mathletters}
\begin{equation}
\frac{\partial {\cal C}_{\;j}^i}{\partial \varphi }=0
\end{equation}

(vi) $C_{\;j}^i$ can be put in the diagonal form $C_{\;j}^i(t,z)=diag$ $%
\left( \alpha ,\beta ,\gamma \right) $ by a suitable choice of coordinates $%
z $ and $t,$ where $\alpha ,\beta $ and $\gamma $ are functions of $t$ and $%
z.$

Due to the presence of the line $L$ the energy-momentum tensor $T_\nu ^\mu ,$
which describes the total content of matter distributed over the whole
space-time, may be decomposed into two parts: $T_{\;\nu }^\mu =\left(
T_{reg}\right) _{\;\nu }^\mu +{\cal L}_{\;\nu }^\mu \delta \left( \rho
\right) ,$ where $\left( T_{reg}\right) _{\;\nu }^\mu $ accounts for the
regular matter surrounding the line and ${\cal L}_\nu ^\mu $ describes the
physical properties of the matter concentrated on $L.$

Now, if we multiply Eq. (\ref{bd1}) by $\rho ,$ take the limit $\rho
\rightarrow 0$ and assume that $\left( T_{reg}\right) _{\;\nu }^\mu $
satisfies the normal dominated convergence condition ($\lim_{\rho
\rightarrow 0}\left[ \sqrt{-g}\left( T_{reg}\right) _{\;\nu }^\mu \right] =0$%
), we obtain 
\begin{equation}
\lim_{\rho \rightarrow 0}\rho \frac \partial {\partial \rho }\left( \phi 
{\cal K}_j^i\right) =0.
\end{equation}

Therefore for small values of $\rho $ we have asymptotically $\phi {\cal K}%
_j^i\simeq {\cal C}_j^i+{\cal O}_j^i,$ where the order of ${\cal O}_j^i$ is
greater than that of $\rho ^\delta ,$ with $\delta >0.$ It follows that 
\begin{equation}
\lim_{\rho \rightarrow 0}\phi {\cal K}_j^i={\cal C}_j^i,  \label{C}
\end{equation}
where it must be remembered that ${\cal C}_j^i$ does not depend on $\varphi $
by virtue of condition (v).

Likewise, Eq. (\ref{bd4}) yields 
\begin{equation}
\lim_{\rho \rightarrow 0}\left( \sqrt{-g}\phi _{,\rho }\right) =\ell (t,z),
\label{l}
\end{equation}
where for consistency we are assuming that $\frac{\partial \ell }{\partial
\varphi }=0.$

It is worth mentioning that the existence of these limits determines the
asymptotic equations (when $\rho \rightarrow 0$) for the metric and the
scalar field. Indeed, from the definition of extrinsic curvature we have in
Gaussian coordinates ${\cal K}_j^i=\frac 12\sqrt{-g}g^{ik}g_{kj,\rho }$ ;
whence from (\ref{C}) for $\rho \rightarrow 0$ it follows that 
\begin{equation}
\frac 12\phi \sqrt{-g}g^{ik}g_{kj,\rho }={\cal C}_j^i.  \label{eq-asym}
\end{equation}
Taking the trace of this equation we get 
\begin{equation}
\phi \frac{\partial \sqrt{-g}}{\partial \rho }={\cal C}.
\label{eq-asym-trace}
\end{equation}

Clearly, the asymptotic behavior of the term $\phi \sqrt{-g}$ is obtained by
taking together (\ref{l}) and (\ref{eq-asym-trace}). Thus, assuming that $%
\lim_{\rho \rightarrow 0}\phi \sqrt{-g}=0$ we are led to the equation 
\begin{equation}
\phi \sqrt{-g}=\left( {\cal C+}\ell \right) \rho ,  \label{phi-asym}
\end{equation}
for $\rho \rightarrow 0.$ Now, substituting (\ref{phi-asym}) into (\ref
{eq-asym}) and considering that ${\cal C}_j^i$ has no degenerate eigenvalue
we obtain, for $\rho \rightarrow 0,$%
\begin{equation}
g_{ij}=diag\left( h_{tt}(t,z)\rho ^{2a},h_{\varphi \varphi }(t,z)\rho
^{2b},h_{zz}(t,z)\rho ^{2c}\right) ,  \label{metric-asym}
\end{equation}
where $h_{tt},h_{\varphi \varphi },$ and $h_{zz}$ are arbitrary functions of 
$t$ and $z,$ and $a\left( t,z\right) ,$ $b\left( t,z\right) $ and $c\left(
t,z\right) $ are defined by the equations $a=\frac{{\cal C}_t^t}{{\cal C}%
+\ell },$ $b=\frac{{\cal C}_\varphi ^\varphi }{{\cal C}+\ell }$ and $c=\frac{%
{\cal C}_z^z}{{\cal C}+\ell }.$ Let us note that in the degenerate case $%
{\cal C}_t^t={\cal C}_z^z$ by transforming the coordinates $t$ and $z$ it is
still possible to put the metric in the form (\ref{metric-asym}).

As to the scalar field, from (\ref{metric-asym}) we have $\sqrt{-g}=\sqrt{-h}%
\rho ^{\frac{{\cal C}}{{\cal C}+\ell }},$ with $h=h_{tt}h_{\varphi \varphi
}h_{zz}.$ Then, from (\ref{phi-asym}) we find directly that for $\rho
\rightarrow 0$ we have 
\begin{equation}
\phi \left( \rho ,t,z\right) =\psi \left( z,t\right) \rho ^d,
\end{equation}
with $d\equiv \frac \ell {{\cal C}+\ell }$ and the function $\psi \left(
z,t\right) $ is defined by $\psi \left( z,t\right) \sqrt{-h}={\cal C}+\ell .$

Now, if we multiply the field equation (\ref{bd3}) by $\left( \phi \sqrt{-g}%
\right) ^2$ and then take the limit $\rho \rightarrow 0$ we readily obtain 
\begin{equation}
{\cal C}_j^i{\cal C}_i^j-{\cal C}^2+\omega \ell ^2-2{\cal C}\ell =0.
\end{equation}

In terms of the functions $a,$ $b,$ $c$ and $d$ we have following
constraints: 
\begin{equation}
a+b+c+d=1  \label{abcd}
\end{equation}
and 
\begin{equation}
a^2+b^2+c^2+(1+\omega )d^2=1.  \label{abcd2}
\end{equation}
Let us note that if $d\rightarrow 0$ more quickly than $\omega ^{-\frac 12}$
when $\omega \rightarrow \infty ,$ then in this limit the General
Relativistic case is recovered \cite{israel}.

These results give a description of the space-time geometry in the exterior
region near the simple line. However, to characterize completely the
energy-momentum tensor of the line we need to consider the ``interior'' of
the line, i.e., the matter tubes whose idealization is pictured by the line.

\section{ Models of matter in the interior of the line}

It seems reasonable to define the energy-momentum tensor of a simple line
source by the expression 
\begin{equation}
{\cal L}_{\;\nu }^\mu =\lim_{\varepsilon \rightarrow 0}\int_0^\varepsilon
\int_0^{2\pi }T_{\;\nu }^\mu \sqrt{-g}d\rho d\varphi .
\end{equation}

It is worth mentioning, however, that this definition may not be useful to
any kind of matter distribution concentrated along the line. In other words,
the matter distribution should satisfy some conditions. For example, the
exterior space-time generated by the matter distribution must be compatible
with the geometry of simple lines discussed in the previous section, whereas
in the interior of the tube matter must possess some special physical
properties.

The first condition to be required refers to ``axial symmetry''. Thus, in
the interior of the tube we impose $\frac{\partial T_{\;\nu }^\mu }{\partial
\varphi }=0;$ whence it must follow that $\frac{\partial \phi }{\partial
\varphi }=0.$ A second condition on the energy-momentum tensor requires that
the radial pressure component $T_{\;\rho }^\rho $ be much less than the
other components of $T_{\;\nu }^\mu .$ Further, let us admit that there
exists a coordinate system in which the interior metric has the form 
\begin{equation}
ds^2=d\rho ^{\prime \,2}+g_{mn}dx^mdx^n+g_{\varphi \varphi }d\varphi ^2,
\end{equation}
where $x^m=(t,z),$ $0\leq \rho ^{\prime }\leq \varepsilon $ and $\frac{%
\partial g_{ij}}{\partial \varphi }=0$ (due to the fact that $\frac{\partial
T_\nu ^\mu }{\partial \varphi }=0).$ It is important to note that we are
assuming that the internal coordinates $\left( t,z,\varphi \right) $ may be
taken continuous at $\rho ^{\prime }=\varepsilon $ by Lie-transporting the
exterior coordinates inward along radial geodesics \cite{israel}. On the
other hand, $\rho $ (the external radial coordinate) and $\rho ^{\prime }$
may not be continuous on $\rho ^{\prime }=\varepsilon $, i.e., at the
boundary of tube these coordinates may have different values. While $\rho
^{\prime }=\varepsilon ,$ a relation $\rho =\rho \left( \varepsilon \right) $
may be obtained from the continuity of the metric.

If we assume that the metric is regular near the tube axis $L,$ then we must
have $g_{\varphi \varphi }=\rho ^{\prime \,2}$ when $\rho ^{\prime
}\rightarrow 0.$ Moreover, the bidimensional metric $g_{mn}$ must be
invertible, that is, $\sqrt{-\;^{\left( 2\right) }g}\neq 0$ at $\rho
^{\prime }=0.$ As to Brans-Dicke scalar field it is natural to require that $%
\phi $ is smooth on the axis, hence we assume that $\lim_{\rho ^{\prime
}\rightarrow 0}\left( \phi _{int}\right) =\eta \left( t,z\right) \neq 0.$
Further assumptions are related to convergence and are given by $%
\lim_{\varepsilon \rightarrow 0}\left( \varepsilon \left. \phi \sqrt{-g}%
^{\left( 3\right) }R_j^i\right| _\xi \right) =0,$ $\lim_{\varepsilon
\rightarrow 0}\left( \varepsilon \left. \phi \sqrt{-g}^{\left( 3\right)
}W_j^i\right| _\xi \right) =0$ and $\lim_{\varepsilon \rightarrow 0}\left(
\varepsilon \left. \sqrt{-g}^{\left( 3\right) }\nabla ^2\phi _{int}\right|
_\xi \right) =0,$ for every $\xi $ inside the interval $0\leq \xi \leq
\varepsilon .$ Then, taking into account these conditions and integrating
the field equations (\ref{bd1}) and (\ref{bd4}) with respect to $\rho ,$ we
obtain 
\begin{mathletters}
\label{bdfinal}
\begin{eqnarray}
\lim_{\varepsilon \rightarrow 0}\left. \phi {\cal K}_j^i\right|
_0^\varepsilon &=&-4\left( {\frak L}_j^i-\frac 12\delta _j^if(\omega ){\frak %
L}\right)  \label{bdlinha} \\
\lim_{\varepsilon \rightarrow 0}\left. \sqrt{-g}\phi ,_{\rho ^{\prime
}}\right| _0^\varepsilon &=&\frac 4{3+2\omega }{\frak L}  \label{bdphi}
\end{eqnarray}

The terms evaluated at $\rho ^{\prime }=\varepsilon $ can be computed by
using the exterior solution since we are assuming continuous junction
between the external and internal solutions. However, for $\rho ^{\prime }=0$
these terms depend upon the internal solution and cannot be determined
exactly, hence must be estimated somehow. In fact, such estimates are
possible by virtue of some conditions we have assumed previously.

The components of the extrinsic curvature $K_j^i$ which depend on $t$ and $z$
only are finite quantities due to the fact that the axis $L$ is, by
assumption, a smooth surface. Then, since $\phi \sqrt{-g}\rightarrow \eta 
\sqrt{-\,^{\left( 2\right) }g}\rho ^{\prime },$ for $\rho ^{\prime
}\rightarrow 0,$ we have 
\end{mathletters}
\begin{equation}
\left. \phi {\cal K}_n^m\right| _0=0,  \label{Ktz}
\end{equation}
with $m,n=t,z.$ To calculate the term $\left. \phi {\cal K}_\varphi ^\varphi
\right| _0$ we just note that in the interior of the tube we have 
\begin{equation}
\left. \phi {\cal K}_\varphi ^\varphi \right| _0=\left. \eta \sqrt{%
-\,^{\left( 2\right) }g}\right| _0.  \label{Kphiphi-general}
\end{equation}

On the other hand, from (\ref{bd3}) we can derive the following equation: 
\begin{equation}
K_\varphi ^\varphi =-\frac 12\frac{\left[ 2\det (K_n^m)+K_m^m\frac{\phi
,_{\rho ^{\prime }}}\phi -\frac 12\omega \left( \frac{\phi ,_{\rho ^{\prime
}}}\phi \right) ^2+^{\left( 3\right) }R+^{\left( 3\right) }W+16\pi T_{\rho
^{\prime }}^{\rho ^{\prime }}\right] }{K_m^m+\left( \frac{\phi ,_{\rho
^{\prime }}}\phi \right) }  \label{bd-vinc}
\end{equation}
Since the cylinders $\rho ^{\prime }=const$ are regular hypersurfaces, then,
if the numerator of equation (\ref{bd-vinc}) does not vanish, we can
conclude that $K_m^m+\frac{\phi ,_{\rho ^{\prime }}}\phi \neq 0$ for $0<\rho
^{\prime }<\varepsilon .$ By virtue of the continuous junction between the
external and internal solutions at $\rho ^{\prime }=\varepsilon ,$ we have 
\begin{equation}
\left[ K_m^m+\left( \frac{\phi ,_{\rho ^{\prime }}}\phi \right) \right]
_{int}=\left[ K_m^m+\left( \frac{\phi ,_{\rho ^{\prime }}}\phi \right)
\right] _{ext}=\frac{1-b}{\rho \left( \varepsilon \right) }.
\end{equation}
Thus, we are left with three cases to be investigated: $0<b<1,$ $b=1$ and $%
\dot{b}>1$. As we shall see later, the case $b>1$ represents a distinctive
feature of Brans-Dicke theory.

a) Case $0<b<1.$ In this case we have $K_m^m+\left( \frac{\phi ,_{\rho
^{\prime }}}\phi \right) >0$ at the boundary of the tube. Since this term
does not vanish in the interior region, it does not change its sign. Hence,
we conclude that $K_m^m+\left( \frac{\phi ,_{\rho ^{\prime }}}\phi \right)
>0 $ in the interval $0<\rho ^{\prime }<\varepsilon .$ On the other hand, it
is easily shown that $K_m^m+\left( \frac{\phi ,_{\rho ^{\prime }}}\phi
\right) =\frac \partial {\partial \rho ^{\prime }}\ln \left( \left| \phi
\right| \sqrt{-\;^{(2)}g}\right) .$ Therefore, for $b<1$ the function $%
\left| \phi \right| \sqrt{-\;^{(2)}g}$ increases monotonically in the
interval $0<\rho ^{\prime }<\varepsilon .$ Thus, we have 
\begin{equation}
\left| \phi {\cal K}_\varphi ^\varphi \right| _0=\left( \left| \phi \right| 
\sqrt{-\;^{(2)}g}\right) _0<\left( \left| \phi \right| \sqrt{-\;^{(2)}g}%
\right) _\varepsilon =\left| \psi \right| \sqrt{-\;^{(2)}h}\left[ \rho
(\varepsilon )\right] ^{1-b},
\end{equation}
where $\sqrt{-\;^{(2)}h}=\sqrt{-h_{tt}h_{zz}}.$ Since $\rho \left(
\varepsilon \right) \rightarrow 0$ when $\varepsilon \rightarrow 0,$ we
conclude that 
\begin{equation}
\lim_{\varepsilon \rightarrow 0}\left| \phi {\cal K}_\varphi ^\varphi
\right| _0=0.  \label{Kphiphi-(b<1)}
\end{equation}

b) Case $b>1.$ Here the situation is exactly opposite to the former case.
Since $K_m^m+\left( \frac{\phi ,_{\rho ^{\prime }}}\phi \right) <0$ the
function $\left| \phi \right| \sqrt{^{(2)}g}$ decreases monotonically for $%
0<\rho ^{\prime }<\varepsilon .$ Then, 
\begin{equation}
\left| \phi {\cal K}_\varphi ^\varphi \right| _0=\left( \left| \phi \right| 
\sqrt{^{(2)}g}\right) _0>\left( \left| \phi \right| \sqrt{^{(2)}g}\right)
_\varepsilon =\left| \psi \right| \sqrt{^{(2)}h}\left[ \rho (\varepsilon
)\right] ^{1-b}.
\end{equation}
Therefore, 
\begin{equation}
\lim_{\varepsilon \rightarrow 0}\left. \phi {\cal K}_\varphi ^\varphi
\right| _0\rightarrow \infty .
\end{equation}
Clearly, this case is unphysical and should be discarded, so $b$ must be
restricted to the interval $0<b\leq 1.$

c) Case $b=1.$ We cannot apply the previous analysis when $b=1.$ Thus, let
us use the following argument. Consider the interior metric and the scalar
field written respectively in the form 
\begin{mathletters}
\begin{eqnarray}
ds^2 &=&d\rho ^{\prime \text{ }2}+\left[ M_\varepsilon (t,z)+m_\varepsilon
(\rho ^{\prime },t,z)\right] dt^2+\left[ N_\varepsilon (t,z)+n_\varepsilon
(\rho ^{\prime },t,z)\right] dz^2+  \nonumber \\
&&\left[ L_\varepsilon (t,z)+l_\varepsilon (\rho ^{\prime },t,z)\right]
dtdz+P_\varepsilon (\rho ^{\prime },t,z)d\varphi ^2, \\
\phi (\rho ^{\prime },t,z) &=&\eta _\varepsilon (t,z)+\sigma _\varepsilon
(\rho ^{\prime },t,z),
\end{eqnarray}
where, by regularity conditions, we must have $m_\varepsilon
(0,t,z)=n_\varepsilon (0,t,z)=l_\varepsilon (0,t,z)=\sigma _\varepsilon
(0,t,z)=0$ and $P_\varepsilon (\rho ^{\prime },t,z)\simeq \rho ^{\prime 
\text{ }2}$ in the limit $\rho ^{\prime }\rightarrow 0.$ By virtue of
continuity at $\rho ^{\prime }=\varepsilon $ we have the relations 
\end{mathletters}
\begin{mathletters}
\begin{eqnarray}
M_\varepsilon (t,z)+m_\varepsilon (\varepsilon ,t,z) &=&h_{tt}(t,z)\rho
(\varepsilon )^{2a} \\
N_\varepsilon (t,z)+n_\varepsilon (\varepsilon ,t,z) &=&h_{zz}(t,z)\rho
(\varepsilon )^{2c} \\
L_\varepsilon (t,z)+l_\varepsilon (\varepsilon ,t,z) &=&0 \\
\eta _\varepsilon (t,z)+\sigma _\varepsilon (\varepsilon ,t,z) &=&\psi
(t,z)\rho (\varepsilon )^d
\end{eqnarray}
Then, we have

\end{mathletters}
\begin{eqnarray}
\left. \phi {\cal K}_\varphi ^\varphi \right| _{\rho ^{\prime }=0}=\left(
\psi \rho (\varepsilon )^d-\sigma _\varepsilon (\varepsilon ,t,z)\right)
&&[\left( h_{tt}\rho (\varepsilon )^{2a}-m_\varepsilon (\varepsilon
,t,z)\right) \left( h_{zz}\rho (\varepsilon )^{2c}-n_\varepsilon
(\varepsilon ,t,z)\right)  \nonumber \\
&&-\frac 14\left( l_\varepsilon (\varepsilon ,t,z)\right) ^2]^{\frac 12}.
\end{eqnarray}
If we do not want the result above to depend explicitly on the details of
the internal solution, then we must restrict ourselves to models of matter
which $m_\varepsilon ,n_\varepsilon ,l_\varepsilon $ and $\sigma
_\varepsilon $ go to zero as $\varepsilon \rightarrow 0$ and satisfy the
following convergence requirements: 
\begin{equation}
\lim_{\varepsilon \rightarrow 0}\rho ^{-2a}m_\varepsilon (\varepsilon
,t,z)=\lim_{\varepsilon \rightarrow 0}\rho ^{-2c}n_\varepsilon (\varepsilon
,t,z)=\lim_{\varepsilon \rightarrow 0}\rho ^dl_\varepsilon (\varepsilon
,t,z)=\lim_{\varepsilon \rightarrow 0}\rho ^{-d}\sigma _\varepsilon
(\varepsilon ,t,z)=0.  \label{mnl}
\end{equation}
Therefore, in this case 
\begin{equation}
\lim_{\varepsilon \rightarrow 0}\left. \phi {\cal K}_\varphi ^\varphi
\right| _0=\psi \sqrt{-\;^{(2)}h}.  \label{kphiphi}
\end{equation}

Let us now look into the equation (\ref{bdphi}). We expect, by regularity
requirements, that the scalar field $\phi $ is finite in the axis $L$ of the
matter tube. It follows that $\lim_{\rho ^{\prime }\rightarrow 0}\phi _{int}$
is finite. Whence we have $\lim_{\rho ^{\prime }\rightarrow 0}\left( \rho
^{\prime }\phi _{int,\rho ^{\prime }}\right) =0.$ Then, since $\sqrt{-g}%
\simeq \sqrt{^{\left( 2\right) }g}\rho ^{\prime }$ near the tube axis, we
conclude that 
\begin{equation}
\left. \sqrt{-g}\phi ,_{\rho ^{\prime }}\right| _0=0,  \label{l-interior}
\end{equation}
a result that holds for any value of $b.$

Taking the above results into account we now are able to find out the field
equations for a simple line. The cases $0<b<1$ and $b=1$ must be considered
separately.

a) Case $0<b<1.$ The equations (\ref{bdlinha}) and (\ref{bdphi}) yield 
\begin{mathletters}
\label{b<1}
\begin{eqnarray}
{\frak L}_{\;j}^i &=&-\frac 14\left[ {\cal C}_{\;j}^i-\delta _{\;j}^i\left( 
\frac{1+\omega }\omega \right) {\cal C}\right] ,  \label{L-b<1} \\
\ell &=&\frac 4{2\omega +3}{\frak L.}  \label{phi-b<1}
\end{eqnarray}
The latter equation may be read as $\ell =\frac 1\omega {\cal C}$ and
represents a further constraint on the functions $a,$ $b,$ $c$ and $d.$
Hence, we have the following set of constraints: 
\end{mathletters}
\begin{mathletters}
\label{acd-b<1}
\begin{eqnarray}
d &=&\frac 1{1+\omega } \\
a+b+c &=&\frac \omega {1+\omega } \\
a^2+b^2+c^2 &=&\frac \omega {1+\omega }
\end{eqnarray}
It turns out that the above system of algebraic equations can have real
solutions only if $\omega <-\frac 32$ or $\omega >0.$

b) Case $b=1.$ From (\ref{Ktz}) e (\ref{kphiphi}), we can write the
equations (\ref{bdlinha}) e (\ref{bdphi}) in the form 
\end{mathletters}
\begin{mathletters}
\label{b=1}
\begin{eqnarray}
{\frak L}_j^i &=&-\frac 14\left( \left[ {\cal C}_j^i\right] -\delta
_j^i\left( \frac{1+\omega }\omega \right) \left[ {\cal C}\right] \right)
\label{L-b=1} \\
\ell &=&\frac 4{2\omega +3}{\frak L}=\frac 1\omega \left[ {\cal C}\right] ,
\label{phi-b=1}
\end{eqnarray}
where $\left[ {\cal C}_{\;j}^i\right] ={\cal C}_{\;j}^i-\psi \sqrt{%
-\;^{\left( 2\right) }h}\delta _\varphi ^i\delta _j^\varphi .$ From the
asymptotic form of the metric given by (\ref{metric-asym}) a straightforward
calculation yields ${\cal C}=$ $\psi \sqrt{-\;^{\left( 2\right) }h}\lambda
(1-d),$ where we are defining $\lambda \equiv \sqrt{h_{33}}.$ Now, from the
equation $\ell =\frac 1\omega ({\cal C}$ $-$ $\psi \sqrt{-\;^{\left(
2\right) }h})$ and taking into account the constraints (\ref{abcd}) and (\ref
{abcd2}) we have for $b=1,$ the following system of equations: 
\end{mathletters}
\begin{mathletters}
\label{acd-b=1}
\begin{eqnarray}
a+c &=&-\frac 1{1+\omega }q \\
a^2+c^2 &=&-\frac 1{1+\omega }q^2 \\
d &=&\frac 1{1+\omega }q
\end{eqnarray}
with $q\equiv (1-\frac 1\lambda ).$

This completes our analysis. Equations (\ref{L-b<1}), (\ref{phi-b<1}),(\ref
{L-b=1}) and (\ref{phi-b=1}) together with the constraints relations
represent in the context of Brans-Dicke the field equations describing the
simple line $L.$ In the next section we apply this formulation to the case
of a static space-time with cylindrical symmetry.

\section{Simple line sources in the static space-times with cylindrical
symmetry}

It is known that the general form of the metric of a static space-time with
cylindrical symmetry may be written in the form \cite{hiscock} 
\end{mathletters}
\begin{equation}
ds^2=d\rho ^2-e^{A(\rho )}dt^2+e^{B(\rho )}d\varphi ^2+e^{C(\rho )}dz^2,
\label{metricacilindrica}
\end{equation}
where $-\infty <t,z<\infty ,$ $\rho >0$ e $0<\varphi <2\pi .$ For the scalar
field we must have $\phi =\phi (\rho )$ by virtue of symmetry.

Let us consider the space-time generated by a singular source concentrated
along the axis $\rho =0.$ Thus, for $\rho \neq 0$ the metric and the scalar
field must satisfy Brans-Dicke field equations (\ref{bd}) in vacuum. The
general solution for the metric and the scalar field is given by 
\begin{mathletters}
\label{vacuum}
\begin{eqnarray}
ds^2 &=&d\rho ^2-\rho ^{2a}dt^2+\lambda ^2\rho ^{2b}d\varphi ^2+\rho
^{2c}dz^2  \label{metricacilindricavacuo} \\
\phi  &=&\phi _0\rho ^d,  \label{phivacuo}
\end{eqnarray}
where $a,$ $b,$ $c,$ $d$ and $\phi _0$ are constants that must satisfy the
constraint equations (\ref{abcd}) and (\ref{abcd2}).

It is reasonable assuming that, in the limit $\omega >>1,$ $\phi _0$ has the
limit \cite{brans}: 
\end{mathletters}
\begin{equation}
\lim_{\omega \rightarrow \infty }\phi _0=\frac 1G,
\end{equation}
with $G$ denoting the gravitational constant.

We see that (\ref{metricacilindricavacuo}) coincides with the asymptotic
form of the metric generated by a simple line when $\rho \rightarrow 0.$
Therefore, to determine the energy-momentum tensor associated with the
source which generates the space-time (\ref{metricacilindricavacuo}), two
cases must be investigated separately: $0<b<1$ or $b=1.$

a) Case $0<b<1.$ In this case it is easy to see that (\ref{L-b<1}) and (\ref
{phi-b<1}) yield 
\begin{mathletters}
\begin{eqnarray}
{\frak L}_{\;j}^i &=&\frac 14\phi _0\lambda diag(1-a,1-b,1-c) \\
\phi  &=&\phi _0\rho ^{\frac 1{1+\omega }},
\end{eqnarray}
with the constants $a,$ $b$ and $c$ satisfying the constraint relations (\ref
{acd-b<1}).

At this point two remarks should be made. One refers to the fact that if $%
\left| \omega \right| \rightarrow \infty ,$ then the solution (\ref
{metricacilindricavacuo}) reduces to the General Relativity solution
corresponding to Kasner vacuum metric which represents the gravitational
field of an infinite rod \cite{israel}. The second remark concerns the
possibility of the linear mass density of the source $\mu \equiv -{\frak L}%
_t^t$ being positive as it can happen for $\omega <-2.$

b) Case $b=1.$ From the equations (\ref{b=1}) and setting $b=1$ in (\ref
{metricacilindricavacuo}), it follows: 
\end{mathletters}
\begin{equation}
\left[ {\cal C}_{\;j}^i\right] =\phi _0\lambda diag(a,q,c),
\end{equation}
Hence, 
\begin{equation}
{\frak L}_{\;j}^i=-\frac 14\phi _0\lambda diag(a-q,0,c-q).  \label{L-q}
\end{equation}
As we have mentioned earlier, in the case $b=1,$ the constants $a$ e $c$
must satisfy (\ref{acd-b=1}). Then, we can express $a$ and $c$ in terms of $%
q:$%
\begin{mathletters}
\label{acq}
\begin{eqnarray}
a &=&-\frac{1/2}{1+\omega }\left( 1\pm \sqrt{-\left( 2\omega +3\right) }%
\right) q \\
c &=&-\frac{1/2}{1+\omega }\left( 1\mp \sqrt{-\left( 2\omega +3\right) }%
\right) q \\
d &=&\frac 1{1+\omega }q
\end{eqnarray}
where we must impose $\omega <-\frac 32,$ otherwise these constants will not
be real numbers. Substituting (\ref{acq}) into (\ref{L-q}) and recalling
that $\lambda q=\lambda -1$ we obtain 
\end{mathletters}
\begin{equation}
{\frak L}_{\;j}^i=-\frac 14(1-\lambda )diag\left( \xi _{\pm }(\omega ),0,\xi
_{\mp }(\omega )\right) ,  \label{L-esquerdo}
\end{equation}
where we define 
\begin{equation}
\xi _{\pm }(\omega )\equiv \frac{\left( 2\omega +3\right) \pm \sqrt{-\left(
2\omega +3\right) }}{2(1+\omega )}\phi _0.
\end{equation}

Thus, from (\ref{L-esquerdo}) we conclude that the energy-momentum tensor
must have the form 
\begin{equation}
{\frak L}_j^i=diag(-\mu ,0,\tau _z),
\end{equation}
with an equation of state given by 
\begin{equation}
\frac{\tau _z}\mu =-\frac{\xi _{\mp }(\omega )}{\xi _{\pm }(\omega )}.
\end{equation}

We can see that the metric induced on the surfaces $t=const$ and $z=const$
coincides with the metric of the cone, with the angular deficit given by the
constant $\lambda .$ From (\ref{L-esquerdo}), a connection between the
energy density of the line $\mu $ and the angular deficit $\lambda $ is
established: 
\begin{equation}
\lambda =1-\frac{4\mu }{\xi _{\pm }(\omega )}.
\end{equation}

It is not difficult to see that in the case of the solution $\xi _{-}(\omega
)$ if we take $-2<\omega <-\frac 32,$ then rather than angular deficit we
have an angular excess, even for a source with positive energy density. This
situation is completely new and peculiar to Brans-Dicke theory in comparison
to the results obtained in General Relativity \cite{israel}.

Finally, it should be mentioned that in the light of the above results the
well-known Gundlach and Ortiz cosmic string \cite{gundlach} is not included
in the set of solutions considered previously. We shall discuss this point
in the next section.

\section{The Gundlach and Ortiz cosmic string}

As we have seen previously the Brans-Dicke field equations for line sources
are given by (\ref{bdfinal}) which by virtue of (\ref{Ktz}), (\ref
{Kphiphi-general}) and (\ref{l-interior}) may be rewritten in the form 
\begin{mathletters}
\label{eq-bdline}
\begin{eqnarray}
C_j^i-\delta _\varphi ^i\delta _j^\varphi \zeta  &=&-4\left( {\cal L}_j^i-%
\frac 12f(\omega ){\cal L}\right)  \\
\ell  &=&\frac 4{3+2\omega }{\cal L}
\end{eqnarray}
with ${\cal C}_j^i$ and $\ell $ given by (\ref{C}) and (\ref{l})
respectively, and 
\end{mathletters}
\begin{equation}
\zeta \equiv \lim_{\varepsilon \rightarrow 0}\left. \eta \sqrt{-\;^{\left(
2\right) }g_{int}}\right| _{\rho ^{\prime }=0}  \label{interiorterm}
\end{equation}
where $g_{int}$ denotes the determinant of $g_{mn}$ in the interior region
and $\eta =\lim_{\rho ^{\prime }\rightarrow 0}\left( \phi _{int}\right) $ as
defined in section III.

It turns out that from (\ref{eq-bdline}) one can relate the geometry and the
scalar field generated in a region close to the line (which are present in
the equations as ${\cal C}_j^i$ and $\ell $) with the physical properties of
the source described by ${\cal L}_j^i.$ However by only knowing ${\cal L}_j^i
$ one cannot from these equations determine the metric and scalar field due
to the presence of $\zeta $ which, as can be seen from its definition (\ref
{interiorterm}), depends on the internal structure of the source. Therefore,
to work out the field equations (\ref{eq-bdline}) some informations
concerning the internal matter distribution are needed, so that $\zeta $ may
be estimated. In this way simple lines would be classified according to the
value of $\zeta $ and, then, models corresponding to distinct classes would
satisfy distinct field equations. In the previous section we have considered
two distinct classes:

1) Case $\zeta =0$ (see eq. (\ref{Kphiphi-(b<1)})). This corresponds to
models for which $b<1$ and the numerator of equation (\ref{bd-vinc}) does
not vanish in the interior region. As we have seen, these assumptions demand
that $\zeta =0.$

2) Case $\zeta =\psi \sqrt{-\;^{\left( 2\right) }h}$ (see eq. (\ref{kphiphi}%
)). Here the models are required to satisfy the convergence conditions given
by (\ref{mnl}).

As far as the Gundlach and Ortiz cosmic string is concerned one can easily
verify that it does not belong to any of the two cases mentioned above. In
fact, one can show that the Gundlach and Ortiz solution corresponds to
choosing $\zeta =\psi \sqrt{-\;^{\left( 2\right) }h}$ with $b<1.$ Clearly,
this choice implies that the condition on the non-vanishing of the numerator
of (\ref{bd-vinc}) must be relaxed.

As already seen, the metric of a static space-time with cylindrical symmetry
and the scalar field which satisfy Brans-Dicke field equation in vacuum are
given by (\ref{vacuum}), where the constants $a$, $b,$ $c$ e $d$ must
satisfy the constraint relations (\ref{abcd}) and (\ref{abcd2}). Now, let us
consider a line source described by ${\cal L}_j^i=\left( -\mu ,p_\varphi
,-\mu \right) .$ We can verify that by choosing $\zeta =\psi \sqrt{%
-\;^{\left( 2\right) }h},$ which reduces to $\zeta =\phi _0$ for solution (%
\ref{vacuum}), since, in this case, $\psi =\phi _0$ and $\sqrt{-\;^{\left(
2\right) }h}=1,$ the field equations (\ref{eq-bdline}) yield 
\begin{mathletters}
\begin{eqnarray}
a &=&c\simeq \frac{4\mu }{3+2\omega }\frac 1{\phi _0} \\
b &\simeq &1-\left( \frac{4\mu }{\phi _0}\right) ^2\frac 1{2+2\omega } \\
d &\simeq &\frac{-8\mu }{3+2\omega }\frac 1{\phi _0} \\
\lambda  &\simeq &1-\frac{4\mu }{\phi _0}\left( \frac{2+2\omega }{3+2\omega }%
\right) 
\end{eqnarray}
where by the symbol $\simeq $ we are denoting the first correction with
respect to linear energy density $\mu .$ Further if the constraint (\ref
{abcd2}) is to be satisfied then one is led to the following additional
equation of state: 
\end{mathletters}
\begin{equation}
p_\varphi \simeq \frac{2\mu ^2}{\phi _0}\frac 1{1+\omega }
\end{equation}

In the limit $\omega >>1$, taken in Gundlach and Ortiz's paper, we expect
that $\phi _0\simeq \frac 1G,$ and so we obtain 
\begin{mathletters}
\begin{eqnarray}
a &=&c\simeq \frac{4\mu G}{3+2\omega } \\
b &\simeq &1-\frac{\left( 4\mu G\right) ^2}{3+2\omega } \\
d &\simeq &\frac{-8\mu G}{3+2\omega } \\
\lambda  &\simeq &1-4\mu G
\end{eqnarray}

At this stage, let us recall some points about the Gundlach and Ortiz's
solution. Firstly, let us note that model assumed for the cosmic string is
described by the action 
\end{mathletters}
\begin{eqnarray}
{\cal S}_{matter} &=&-\int d^4x\sqrt{-g}\left[ \frac 12\left[ (\nabla _\nu
\psi +ieA_\nu )\Phi \right] \left[ (\nabla ^\nu \psi +ieA^\nu )\Phi \right]
^{*}\right.  \nonumber \\
&&\left. \alpha \left( \Phi \Phi ^{*}-\eta ^2\right) ^2+\frac 1{16\pi }%
F_{\nu \kappa }F^{\nu \kappa }\right] ,  \label{action}
\end{eqnarray}
where $A_\nu $ is a vector field, $\Phi $ is a complex scalar field, $\nabla
_\nu $ is the covariant derivative with respect to the space-time metric, $%
F_{\nu \kappa }\equiv \nabla _\nu A_\kappa -\nabla _\kappa A_\nu ,$ and $%
\alpha ,$ $\eta $ and $e$ are constants. Further it is assumed the ansatz 
\begin{eqnarray}
\Phi &=&\eta X\left( \rho \right) \exp \left( i\varphi \right) \\
A_\nu &=&\frac 1e\left( P\left( \rho \right) -1\right) \nabla _\nu \varphi
\end{eqnarray}

Secondly, according to a scheme of approximation the interior solution is
characterized by the fact that effects of the scalar field are small in
comparison with the cosmic string energy-momentum tensor contribution. Thus,
in the interior region, the field equations would be approximately replace
by the Einstein equations. Therefore, one can make use of a result (see \cite
{linet,bogomolnyi}) which states that, in the context of General Relativity,
if one takes $8\alpha =e^2,$ for the lower energy solution, the following
relation is valid 
\begin{equation}
\mu =\pi \eta ^2
\end{equation}

Now, considering this result, we obtain after a coordinate transformation,
the Gundlach and Ortiz's solution in the order of $\eta ^2:$%
\begin{eqnarray}
ds^2 &=&r^{8\pi G\eta ^2\beta ^2}\left[ -dt^2+dz^2+dr^2+\left( 1-4\pi G\eta
^2\right) ^2d\varphi ^2\right]  \\
\phi  &\sim &r^{-8\pi G\eta ^2\beta ^2}
\end{eqnarray}
where $\beta ^2=\frac 1{3+2\omega }$

\section{Thin shells in Brans-Dicke theory}

The case of thin shells or surface layers was also considered by Israel \cite
{israel-sup}, who presented a complete formulation of the problem. In this
section we shall briefly outline an approach to the same problem in
Brans-Dicke theory of gravity.

Let us consider the timelike hypersurface $\Sigma $ which describes the
history of a thin shell of matter in space-time. And let $V$ be a
neighborhood of $\Sigma $ which admits a Gaussian coordinate system. In
terms of these coordinates we can write the metric of space-time as 
\begin{equation}
ds^2=dn^2+g_{ij}\left( n,x^k\right) dx^idx^j,
\end{equation}
where $n$ is the coordinate associated with the vector $\frac \partial {%
\partial n}$ normal to $\Sigma ,$ and $i,j,k=1,2,3.$ We can also choose the
coordinate $n$ such that $n=0$ corresponds to $\Sigma \cap V.$

Similarly to the decomposition (\ref{bd}) we can project Brans-Dicke field
equations (\ref{full-bd}) onto and perpendicular to the hypersurfaces $%
n=const.$ Then, in terms of the extrinsic and intrinsic curvatures of these
hypersurfaces the field equations become identical to the equations (\ref{bd}%
) by just substituting the coordinate $\rho $ for $n,$ with $K_{\;j}^i$ now
denoting the extrinsic curvature of the hypersurface $n=const.$

Let us assume that the hypersurface $\Sigma $ is smooth, with its geometry
described by the tridimensional metric $g_{ij}(n=0,x^k).$ We also suppose
that the scalar field $\phi $ is continuous on $\Sigma ,$ though its
derivative $\phi _{,n}$ may be discontinuous when $\Sigma $ is crossed.

Now, if we integrate the field equations (\ref{bd}) with respect to the
coordinate $n$ in the interval $-\varepsilon <$ $n<\varepsilon ,$ and take
the limit $\varepsilon \rightarrow 0,$ we easily obtain: 
\begin{mathletters}
\label{sup}
\begin{eqnarray}
\left[ K_{\;j}^i\right] &=&-\frac{8\pi }\phi \left( S_{\;j}^i-\frac 12\delta
_{\;j}^if(\omega )S\right)  \label{sup1} \\
0 &=&S_{\;\mu }^n  \label{sup2} \\
\left[ \phi ,_n\right] &=&\frac{8\pi }{3+2\omega }S,  \label{sup3}
\end{eqnarray}
where 
\end{mathletters}
\begin{equation}
S_{\;\nu }^\mu \equiv \lim_{\varepsilon \rightarrow 0}\int_{-\varepsilon
}^\varepsilon T_{\;\nu }^\mu dn,  \label{Sphi}
\end{equation}
defines the energy-momentum tensor of the matter distribution concentrated
on the shell and, as before, the bracket $\left[ X\right] $ of a quantity $X$
denotes the operation 
\begin{equation}
\left[ X\right] \equiv \lim_{\varepsilon \rightarrow 0}\left[ \left.
X\right| _{n=\varepsilon }-\left. X\right| _{n=-\varepsilon }\right] .
\end{equation}

Thus, we see that, as in the case of General Relativity, the equations (\ref
{sup}) relate the physical properties of the matter lying on $\Sigma $ to
the geometry of space-time near $\Sigma .$ From these equations we also see
that when the matter distribution is not regular, and has a singular part
with support on $\Sigma ,$ i.e., $T_{\;\nu }^\mu =\left( T_{reg}\right)
_{\;\nu }^\mu +S_{\;\nu }^\mu \delta \left( n\right) ,$ then the singular
part of the energy-momentum tensor causes a discontinuity to appear in the
extrinsic curvature of $\Sigma .$ In the case of Brans-Dicke theory this
discontinuity depends also on the value the scalar field takes on $\Sigma $
as well as on the parameter $\omega .$ On the other hand, we see that $S_j^i$
also induces a discontinuity on the derivative of the scalar field $\phi
_{,n}$, whose value depends on the trace $S.$

Finally, let us note that when $\omega \rightarrow \infty $ the equations (%
\ref{sup}) become identical to those of General Relativity provided that $%
\phi \rightarrow G$ in this limit \cite{israel-sup}.

\section{Acknowledgments}

The authors wish to thank CNPq (Brazil) for financial support.

\end{document}